\documentclass[12pt,preprint]{aastex}
\slugcomment{draft of \today}
\shorttitle{Re-acceleration of CRs at Cosmological Shocks}
\shortauthors{Kang \& Ryu}

\def\ie{{\it i.e.,~}}
\def\kms{~{\rm km~s^{-1}}}
\def\yrs{~{\rm yrs}}
\def\muG{~{\mu\rm G}}

\begin{document}
\title{Re-acceleration of Nonthermal Particles at Weak Cosmological Shock Waves}

\author{Hyesung Kang$^1$ and Dongsu Ryu$^2$\altaffilmark{,3}}
\affil{$^1$Department of Earth Sciences, Pusan National University, Pusan
609-735, Korea: kang@uju.es.pusan.ac.kr\\
$^2$Department of Astronomy and Space Science, Chungnam National
University, Daejeon, Korea: ryu@canopus.cnu.ac.kr}
\altaffiltext{3}{Author to whom any correspondence should be addressed.}

\begin{abstract}
We examine diffusive shock acceleration (DSA) of the pre-exisiting as
well as freshly injected populations of nonthermal, cosmic-ray (CR)
particles at weak cosmological shocks.
Assuming simple models for thermal leakage injection and Alfv\'enic drift,
we derive analytic, time-dependent solutions for the two populations of
CRs accelerated in the test-particle regime.
We then compare them with the results from kinetic DSA simulations for
shock waves that are expected to form in intracluster media and 
cluster outskirts in the course of large-scale structure formation.
We show that the test-particle solutions provide a good approximation for 
the pressure and spectrum of CRs accelerated at these weak shocks.
Since the injection is extremely inefficient at weak shocks,
the pre-existing CR population dominates over the injected population. 
If the pressure due to pre-existing CR protons is about 5 \% of the
gas thermal pressure in the upstream flow, the downstream CR pressure
can absorb typically a few to 10 \% of the shock ram pressure at shocks
with the Mach number $M \la 3$.
Yet, the re-acceleration of CR electrons can result in a substantial
synchrotron emission behind the shock.
The enhancement in synchrotron radiation across the shock
is estimated to be about a few to several for $M \sim 1.5$ and 
$10^2-10^3$ for $M \sim 3$, depending on the detail model parameters.
The implication of our findings for observed bright radio relics is
discussed.

\end{abstract}


\section{INTRODUCTION}

Cosmological shock waves result from supersonic flow motions induced by
hierarchical clustering during the large-scale structure formation in
the Universe \citep{miniati00, ryuetal03}.
According to studies based on cosmological hydrodynamic simulations,
the shocks formed by merger of subclumps, infall of matter and internal
flow motion in intracluster media (ICMs) and cluster outskirts are
relatively weak with Mach number $M \la$ a few
\citep{ryuetal03, psej06, kangetal07, skillman08, hoeft08, vazza09}.
Indeed, observations of X-ray shocks
\citep[e.g.,][]{markev02, markev05, markev07} and radio relics
\citep[e.g.,][]{bdnp06, fsnw10, wrbh10} indicate that the estimated Mach
number of observed shocks in cluster environments is consistent with 
such theoretical predictions.

Suprathermal particles are known to be produced as an inevitable consequence
of the formation of collisionless shocks in tenuous plasmas and they can be
further accelerated to become cosmic rays (CRs) through
interactions with resonantly scattering Alfv\'en waves in the converging
flow across a shock \citep{bell78, dru83,maldru01}.
Detailed nonlinear treatments of diffusive shock acceleration (DSA) have
predicted that at strong shocks a significant fraction of the shock kinetic
energy is transferred to CRs, inducing highly nonlinear back-reactions
from CRs to the underlying flow \citep[e.g.,][]{ab06,veb06,kj07}.
Multi-band observations of nonthermal radio to $\gamma$-ray emissions
have confirmed the acceleration of CR electrons and protons up to 100 TeV
at young supernova remnants \citep[e.g.][]{parizot06,reynolds08,abdo10}.

The presence of nonthermal particles, especially electrons, in clusters of
galaxies, has been inferred from observations of synchrotron emission from
radio halos and relics \citep[see, e.g.,][for review]{ct02,gf04}.
Since the matter in ICMs and cluster outskirts should have gone first
through accretion shocks of high Mach number around nonlinear structures
and then through weaker shocks due to mergers and flow motion 
\citep{ryuetal03, kangetal07}, DSA should
be responsible for at least a part of the CR production. 
Nonthermal particles can be also produced via turbulent
acceleration \citep[see, e.g.,][]{cb05,bl07}.
Recent Fermi observations of $\gamma$-ray emission from galaxy clusters,
however, limit that the pressure due to CR protons cannot exceed
$\sim 10$ \% of the gas thermal pressure \citep{abdo10,ddcb10}.

At weak shocks with $M \la$ a few, DSA is known to be
rather inefficient and the CR pressure remains dynamically insignificant,
partly because the injection from thermal to nonthermal
particles is inefficient \citep[e.g.,][]{kjg02}.
In such test-particle regime, 
the downstream CR spectrum takes the power-law form of $f_2(p) \propto p^{-q}$,
where the spectral slope, $q$, depends on the velocity jump across the shock
\citep{dru83}.
Recently, \citet{kr10} suggested analytic, time-dependent solutions for
the test-particle CR spectrum, using results from DSA simulations in which
particles are injected via thermal leakage process and accelerated to ever
increasing maximum momentum, $p_{\rm max}(t)$.
They found that at weak shocks expected to form in ICMs and cluster outskirts,
indeed, much less than $\sim 10^{-3}$ of particles are injected into CRs
and much less than $\sim 1 \%$ of the shock ram pressure is converted into the
downstream pressure of CR protons, so the particle acceleration is
virtually negligible.

However, the recent discovery of very bright radio relics associated
with weak shocks of $M \la$ a few \citep[e.g.,][]{bdnp06, fsnw10, wrbh10}
suggests that, contrary to the expectation, DSA should operate 
at weak shocks in cluster environments.
One way to explain this is to presume that the relics form in media with
{\it pre-existing} CRs which were produced by DSA at previous shocks
and/or  by turbulent acceleration.
The existence of pre-exiting CRs alleviates the problem of inefficient
injection at weak shocks.

In this paper, we examine the DSA at weak cosmological shocks in the presence
of pre-existing CRs.
First, the properties of weak shocks in ICMs and cluster outskirts are
briefly reviewed in Section 2.
Analytic, time-dependent solutions for the acceleration of the pre-existing and
freshly injected populations of CRs in the test-particle regime is described
in Section 3, while the numerical solutions from kinetic DSA simulations are 
presented in Section 4.
The synchrotron radiation from CR electrons accelerated at these shocks
is discussed in Section 5.
Finally, a brief summary is given in Section 6.

\section{SHOCK WAVES IN ICMS AND CLUSTER OUTSKIRTS}

Shock waves in the large-scale structure of the universe have been studied in
details using various hydrodynamic simulations for the cold dark matter
cosmology with cosmological constant ($\Lambda$CDM)
\citep{ryuetal03, psej06, kangetal07, skillman08, hoeft08, vazza09}.
It was found that shocks with Mach number typically up to $M \sim 10^3$ and
speed up to $u_s \sim$ a few $\times 1000\ \rm {km\ s}^{-1}$ at the present
universe $(z=0)$.
In ICMs and cluster outskirts, however, shocks are expected to have lower
Mach number, because they form in the hot gas of $kT \ga$ keV.

To examine the characteristics of shocks in ICMs and cluster outskirts,
we analyze the shocks with the preshock gas temperature of $T_1 > 10^7$ K.
The cosmic web is filled with ionized plasmas, the intergalactic medium
 \citep{co99,krcs05}.
The hot gas with $T > 10^7$ K is found mostly in ICMs and cluster outskirts,
and the Warm Hot Intergalactic Medium (WHIM) with $10^5$ K $ < T < 10^7$ K
is distributed mostly in filaments.
The diffuse gas with $T < 10^5$ K resides mainly in sheetlike structures
and voids.
The shocks were found in a simulation of the WMAP1-normalized $\Lambda$CDM
cosmology employed the following parameters: 
$\Omega_{b}=0.048$, $\Omega_{m}=0.31$, $\Omega_{\Lambda}=0.69$,
$h \equiv H_0$/(100 km/s/Mpc) = 0.69, $\sigma_8 = 0.89$, and $n=0.97$. 
The simulation was performed using a PM/Eulerian hydrodynamic
cosmology code \citep{rokc93}.
Detailed descriptions for numerical set-up and input physical ingredients 
can be found in \citet{co06}.
The procedure to identify shocks was described in details in
\citet{ryuetal03}.

Figure 1 shows the surface area of shocks with $T_1 > 10^7$ K per Mach
number interval in the entire simulation volume, normalized by the volume.
Here, $S$ is given in units of $(h^{-1}{\rm Mpc})^{-1}$.
The quantity $S$ provides a measure of shock frequency or the inverse
of the mean comoving distance between shock surfaces.
To avoid confusion from complex flow patterns and shock
surface topologies associated with very weak shocks, only those
portions of shock surfaces with $M\ge1.5$ are shown.
We also calculated the incident shock kinetic energy flux,
$F_{\phi}= (1/2) \rho_1 u_s^3$, where $\rho_1$ is the preshock gas density,
and then the kinetic energy flux through shock surfaces per Mach number
interval, normalized by the simulation volume, $dF_{\phi}(M)/dM$.
Figure 1 shows $dF_{\phi}(M)/dM$, too.
As expected, the Mach number of the shocks formed in ICMs and cluster
outskirts is small, typically $M \la 3$.
The frequency increases to weakest possible shocks with $M \sim 1$.
The kinetic energy flux through shock surfaces is larger for weaker
shocks; that is, weaker shocks process more shock energy, confirming
the energetic dominance of weak shocks in cluster environments.

\section{ANALYTIC TEST-PARTICLE SPECTRUM}

In the kinetic DSA approach, the following diffusion-convection equation 
for the pitch-angle-averaged distribution function of CRs, $f(x,p,t)$, 
is solved along with suitably modified gasdynamic equations:
\begin{equation}
{\partial f \over \partial t}  + (u+u_w) {\partial f \over \partial x}
= {p \over 3} {{\partial (u+u_w)} \over {\partial x}} 
{{\partial f} \over {\partial p}} 
+ {\partial \over \partial x} \left[\kappa(x,p)
{\partial f \over \partial x} \right],
\label{diffcon}
\end{equation}
where $\kappa(x,p)$ is the spatial diffusion coefficient
and $u_w$ is the drift speed of the local Alfv\'enic wave turbulence 
with respect to the plasma \citep{skill75}.
The scattering by Alfv\'en waves tends to isotropize the CR distribution
in the wave frame, which may drift upstream at the Alfv\'en speed, $v_A$,
with respect to the bulk plasma.
So the wave speed is set to be $u_w=-v_A$ upstream of shock, while
$u_w=0$ downstream. 

In the test-particle regime where the feedback due to the CR pressure is
negligible, the downstream CR distribution can be described with a
power-law spectrum, $f_2(p) \propto p^{-q}$, and the slope is given by 
\begin{equation}
q={{3(u_1-v_A)}\over u_1-v_A-u_2}
={{3\sigma(1-M_A^{-1})}\over (\sigma-1-\sigma M_A^{-1})},
\label{qtp}
\end{equation}
where $u_1$ and $u_2$ are the upstream and downstream flow speeds,
respectively, in the shock rest frame, 
$\sigma = u_1/u_2=\rho_2/\rho_1$ is the shock compression ratio, and
$M_A=u_1/v_A$ is the upstream  Alfv\'en Mach number with
$v_A=B_1/\sqrt{4\pi\rho_1}$ \citep{dru83, kr10}.
The test-particle power-law slope $q$ can be calculated as a function of
shock Mach number M with $\sigma = [(\gamma_g+1) M^2]/[(\gamma_g-1) M^2 + 2]$,
which becomes $4 M^2 /(M^2 + 3)$ for a gas adiabatic index $\gamma_g = 5/3$,
and $M_A = M / \delta$.
Here, $\delta \equiv v_A/c_s$ is the Alfv\'en speed parameter, where $c_s$
is the upstream sound speed.
The maximum momentum of CR protons achieved by the shock age of $t$ can be
estimated as
\begin{equation}
p_{\rm max}(t)  \approx m_pc \left[ {{(1-M_A^{-1})
(\sigma -1 -\sigma M_A^{-1})}
\over {3\sigma(2-M_A^{-1})}} \right]
{u_s^2 \over \kappa^*} t ,
\label{pmax}
\end{equation}
where $u_s=u_1$ is the shock speed \citep{dru83, kr10}.
Here, a Bohm-type diffusion coefficient,
\begin{equation}
\kappa(p)= \kappa^* \left( p \over m_p c \right)
\left( \rho_0 \over \rho \right),
\label{diffcoef}
\end{equation}
is adopted, where $\kappa^*= m_pc^3/(3eB_0) = 3.13\times 10^{22}
(B_0/1\muG)^{-1}{\rm cm^2 s^{-1}}$, $B_0$ and $\rho_0$ are magnetic
field strength and the gas density far upstream. 
In CR-modified shocks where CRs are dynamically non-negligible,
in general, the upstream flow is decelerated in  the precursor before it
enters the gas subshock. 
So we use the subscripts ``0", ``1", and ``2" to denote the conditions
far upstream, immediate upstream and downstream of shock,  respectively.
Of course, in the test-particle limit, the distinction between
far and immediate upstream quantities disappears, e.g., $\rho_0=\rho_1$. 

In the limit of large $M$ ($\sigma \approx 4$) and large $M_A$
($\delta \approx 0$), the maximum energy of CR protons
can be approximated by  
\begin{equation}
E_{{\rm max,}p} \approx {{u_s^2 t}\over{8\kappa^*}} m_pc^2
\approx 10^{10}~{\rm GeV} 
\left({u_s \over {10^3 {\rm kms^{-1}}}}\right)^2
\left({t \over {10^9 {\rm yrs}}}\right)
\left({B_0 \over {1\muG}}\right). 
\end{equation}
The CR proton spectrum limited by the shock age is expected to have a cutoff
at around $\sim p_{\rm max}(t)$ (see Section 3.3 for further discussion).

\subsection{Pre-existing Population}

As noted in Introduction, it seems natural to assume that ICMs and
cluster outskirts contain pre-existing CRs.
But their nature is not well constrained, except that $P_c \la 0.1 P_g$, 
\ie the pressure of CR protons is less that $\sim 10$ \% of the gas thermal
pressure \citep[e.g.,][]{abdo10,ddcb10}.
With pre-existing CRs of spectrum $f_0(p)$ upstream of shock, the
steady-state, test-particle solution of Equation (\ref{diffcon})
for the downstream CR distribution can be written as
\begin{equation}
f_2(p) = q p^{-q} \int_{p_{\rm inj}}^p p'^{q-1} f_0(p') dp'
+ f_{\rm inj} \left({p\over p_{\rm inj}}\right)^{-q},
\label{drury}
\end{equation}
where $q$ is the test-particle power-law slope given in Equation (\ref{qtp})
\citep{dru83}.
Here, $p_{\rm inj}$ is the lowest momentum boundary above which particles
can cross the shock, \ie the injection momentum (see the next subsection).
By this definition of $p_{\rm inj}$, the CR distribution function,  
$f_0=0$ and $f_2=0$ for $p<p_{\rm inj}$.
The first term in the right-hand-side of Equation ({\ref{drury}) represents 
the re-accelerated population of pre-existing CRs, while
the second term represents the population of CRs freshly injected 
at the shock and will be  discussed in the next subsection.

We adopt a power-law form,
$f_0(p)=f_{\rm pre}\cdot (p/p_{\rm inj})^{-s}$,
with the slope $s = 4 - 5$, as the model spectrum for pre-existing CR protons. 
If pre-existing CRs were generated at previous shocks, 
the slope of $s = 4 - 5$ is achieved for 
$M \geq \sqrt{5}$ with $\delta = 0$ (see Equation (\ref{qtp})).
On the other hand, if they are mainly the outcome of turbulent acceleration,
the slope should be close to $s \sim 4$ \citep[see, e.g.,][]{chan05}. 
Then, the spectrum of re-accelerated CRs is obtained by direct integration:
\begin{eqnarray}
f_2^{\rm reac}(p) = \left\{ \begin{array}{rl} 
\left[{q/(q-s)}\right] \left[ 1-({p / p_{\rm inj}})^{-q+s}\right] f_0(p),
&\mbox{ if $q\neq s$}\\ 
q \ln(p/p_{\rm inj}) f_0(p), &\mbox{ if $q= s$}.
\end{array} \right.
\label{f2p}
\end{eqnarray}
If $q\ne s$, for $p\gg p_{\rm inj}$,
\begin{equation}
f_2^{\rm reac}(p)= {q \over {|q-s|}} f_{\rm pre}
\left(p \over p_{\rm inj}\right)^{-r},
\label{f2p2}
\end{equation}
where $r=\min(q,s)$.
That is, if the spectral slope of pre-existing CRs is softer than the
test-particle slope ($s>q$), the re-accelerated CR spectrum gets flattened
to $p^{-q}$ by DSA;
in the opposite case ($s<q$), the re-accelerated CR spectrum is simply
amplified by the factor of $q/(q-s)$ and retains the same slope as
the slope of pre-existing CRs.

Figure 2 shows the re-accelerated CR distribution given in Equation
(\ref{f2p}) for a $M=3$ shock in the presence of 
the pre-existing power-law CR spectrum with 
the slope $s= 4$ and  4.5 (right panel) and $s=5$ (left panel). 
The Alfv\'enic drift is ignored ($\delta=0$), 
so the test-particle slope is $q=4.5$.
Here, we adopted the following parameters:
the upstream gas temperature $T_0 = 10^7$ K and
the injection parameter $\epsilon_B = 0.25$,
resulting in $p_{\rm inj} = 8.0 \times 10^{-3} m_p c$
(see the next subsection for details of our injection model).

The figure illustrates that for $p\gg p_{\rm inj}$, the CR amplification
factor, $f_2(p)/f_0(p)$, approaches a constant, $q/(q-s)=9$, in the case
of $s=4$, increases as $\ln(p/p_{\rm inj})$ in the case of $q=s=4.5$,
and scales as $(p/p_{\rm inj})^{s-q}$ in the case of $s=5$. 
So, for instance, the factor becomes $f_2/f_0=32$ and $310$ at $p/m_p c=10$ 
for $s=4.5$ and 5, respectively.
We point that these values of the CR amplification factor are
substantially larger than those expected for the adiabatic compression 
across the shock.
With pre-existing CRs of $f_0 \propto p^{-s}$, the amplification factor
due to the adiabatic compression is given by
\begin{equation}
{f_2^{\rm adb} / f_0} = \sigma^{s/3}
\label{adb-comp}
\end{equation}
in the test-particle regime.
So the adiabatic amplification factor is $f_2^{\rm adb} / f_0 = 4.3$, 5.2, 
and 6.2 and for $s=4$, 4.5 and 5, respectively, at a Mach 3 shock.
Note that the adiabatic compression does not change the slope of the CR
spectrum. 

The left panel of Figure 2 also shows the time evolution of the CR
distribution at the shock location, $f_s(p,t)$, from a DSA simulation for
the same set of parameters (see Section 4 for details of DSA simulations).
The CR injection was turned off for this particular simulation 
in order to compare the analytic and numerical solutions only for
pre-existing CRs. 
This demonstrates that the time-dependent solution asymptotes to the
steady-state solution in Equation (\ref{f2p}).

\subsection{Injected Population}

Because complex plasma interactions among CRs, waves, and the underlying 
gas flow are not fully understood yet, it is not possible to make a precise
quantitative prediction for the injection process from first principles
\citep[e.g.,][]{maldru01}.
Here, we adopt a phenomenological injection model that can emulate
the thermal leakage process, through which particles above a certain
injection momentum $p_{\rm inj}$ cross the shock and get injected to the
CR population \citep{kjg02, kr10}.
Then, the CR distribution function at $p_{\rm inj}$ is anchored to the 
downstream Maxwellian distribution as
\begin{equation}
f_{\rm inj}= f(p_{\rm inj})= {n_2 \over \pi^{1.5}}~ p_{\rm th}^{-3}~
\exp\left(-Q_{\rm inj}^2\right),
\label{finj}
\end{equation}
where $n_2$ is the downstream proton number density. 
Here, $p_{\rm inj}$ and $Q_{\rm inj}$ are defined as 
\begin{equation}
Q_{\rm inj}(M)\equiv {p_{\rm inj} \over p_{\rm th}} \approx 1.17
{{m_p u_2}\over p_{\rm th}} \left(1+ {1.07 \over \epsilon_B}\right)
\left({M\over 3}\right)^{0.1},
\label{Qinj}
\end{equation}
where $p_{\rm th}= \sqrt{2m_p k_B T_2}$ is the thermal peak momentum of
the downstream gas with temperature $T_2$ 
and $k_B$ is the Boltzmann constant.
We note that the functional form of $Q_{\rm inj}$ was adopted to represent
an ``effective'' injection momentum, since particles in the suprathermal
tail can cross the shock with a smoothly-varying probability distribution
\citep[see][]{kjg02}.
One free parameter that controls the leakage process is the injection
parameter, $\epsilon_B = B_0/B_{\perp}$, which is the ratio of the general
magnetic field along the shock normal, $B_0$, to the amplitude of the
downstream, magnetohydrodynamic (MHD) wave turbulence, $B_{\perp}$. 
Although plasma hybrid simulations and theories both suggested that
$0.25 \la \epsilon_B \la 0.35$ \citep{mv98}, the physical range of
this parameter remains to be rather uncertain due to lack of
full understanding of relevant plasma interactions.

The second term in Equation (\ref{drury}) is fixed by $q$, $p_{\rm inj}$,
and $f_{\rm inj}$. 
The fraction of particles injected into the CR population can be estimated
analytically as well:
\begin{equation}
\xi \equiv {n_{CR} \over n_2 } = {4 \over \sqrt{\pi}} Q_{\rm inj}^3
\exp\left(-Q_{\rm inj}^2\right) {1 \over {q- 3}},
\end{equation}
which is fixed only by $Q_{\rm inj}$ and $q$.
The injection fraction depends strongly on
$\epsilon_B$ (through $Q_{\rm inj}$) for weak shocks with $M \la 5$
\citep[see also][]{kr10}. 
For example, it varies from $5\times10^{-5}$ to $10^{-3}$ for
$\epsilon_B=0.25-0.3$ for shocks with $M=3$. 

\subsection{Cosmic-Ray Spectrum for Weak Shocks}

\citet{kr10} demonstrated that the {\it time-dependent}, test-particle
solutions of the downstream CR distribution can be represented by 
the steady-state, test-particle solutions with an exponential cutoff
\citep{capri09}, if the cutoff momentum is set as
$p^*\approx 1.2\ p_{\rm max}(t)$ with $p_{\rm max}(t)$ in Equation
(\ref{pmax}).
Here, we suggest that the same cutoff would be applied to the spectrum of
re-accelerated CRs. 
Then, the CR distribution at the shock location, $x_s$, originated from both
the pre-existing and freshly injected populations can be approximated by 
\begin{equation}
f_s(p,t) \equiv f_2(x_s,p,t) \approx \left[f_2^{\rm reac}(p) + f_{\rm inj}
\cdot \left({p \over p_{\rm inj} }\right)^{-q} \right ] \cdot
\exp\left[- q C(z) \right],
\label{ftest}
\end{equation}
where $f_2^{\rm reac}(p)$ is given in Equation (\ref{f2p}) and $z=p/p^*$.
The function $C(z)$ is defined as 
\begin{equation}
C(z) = \int_{z_{\rm inj}}^z {dz' \over z'} {1\over {\exp(1/z')-1} },
\label{cofz}
\end{equation}
where $z_{\rm inj} = p_{\rm inj}/p^*$ \citep{kr10}.
Of course, for $p > p^*$, the acceleration is limited by the shock age
and so pre-existing CRs will be simply advected downstream,
resulting in $f_s(p)\approx f_0(p)$.
These particles, however, do not make any significant contribution
to the downstream CR pressure, if the pre-existing power-law spectrum 
has the slope $s>4$ (see below).

\section{COMPARISON WITH NUMERICAL SOLUTIONS}

\subsection{Set-up for DSA Simulations}

We carried out kinetic DSA simulations in order to test
the time-dependent features of the test-particle solution in Equation 
(\ref{ftest}).
Also for shocks with typically $M\ga$ a few, the evolution of CR-modified 
shocks should be followed by DSA simulations, 
because the nonlinear feedback of CRs becomes important \citep{kr10}.
We used the CRASH (Cosmic-Ray Acceleration SHock) code for quasi-parallel
shock, in which the diffusion-convection equation (\ref{diffcon}) 
is solved along with the gasdynamic equation modified for the effects 
of the CR pressure \citep{kjg02}.

We considered shocks with a wide range of Mach number, $M=1.5-5$,
propagating into typical ICMs and cluster outskirts of $T_0=10^7$ K; 
the shock speed is $u_s=M \cdot 474\kms$.
The diffusion in Equation (\ref{diffcoef}) was used.
In the code units, the diffusion coefficient is normalized with
$\kappa_o=10^3\kappa^*$ for numerical simulations.
Then, the length and time scales are given as $l_o=\kappa_o/u_s$ 
and $t_o=\kappa_o/u_s^2$, respectively. 
Since the flow structure and $P_c$ profile evolve self-similarly,
a specific physical value of $\kappa_o$ matters only in the determination
of $p_{\rm max}$ at a given simulation time. 
For instance, $p_{\rm max}/m_p c \approx 10^3$ is achieved by the termination
time of $t/t_o=10$ in our simulations.
Simulations start with purely gasdynamic shocks initially at rest
at $x_s = 0$, and the gas adiabatic index is $\gamma_g = 5/3$.

As for the pre-existing CRs, we adopted
$f_0(p) = f_{\rm pre} (p/p_{\rm inj})^{-s}$ for their spectrum. 
The amplitude, $f_{\rm pre}$, is set by the ratio of the upstream CR to gas
pressure, $R\equiv P_{c,0}/P_{g,0}$, and we consider $R = 0.01 - 0.1$.
We note that with the same value of $R$, the amplitude $f_{\rm pre}$ is
larger for softer pre-existing spectrum, \ie larger $s$.
To examine the effects of Alfv\'enic drift, 
in addition to the models with $\delta = 0$,
we consider $\delta = 0.42$ as a fiducial value, which corresponds
to $E_B\sim 0.1 E_g$, \ie the magnetic field energy density of
$\sim 10$ \% of the gas thermal energy density. 
Finally, we consider $\epsilon_B=0.25-0.3$ for the injection parameter.

\subsection{CR Proton Spectrum and CR Pressure}

Figure 3 shows the CR pressure profile and the CR distribution at the shock
location, $f_s$, from DSA simulations for a Mach 3 shock.
In the cases with pre-existing CRs in (b) and (c), the steady-state solution
without injection given in Equation (\ref{f2p}) (dot-dashed line)
is also shown for comparison.
As CRs are accelerated to ever high energies ($p_{\rm max} \propto t$), 
the scale length of the CR pressure increases linearly with time, 
$l_d(p_{\rm max}) \propto u_s t$ \citep{krj09}.
Left panels demonstrate that the CR pressure profile evolves in a
self-similar fashion, depending approximately only on the similarity variable,
$x/(u_s t)$.
Right panels indicate that $f_s$ can be well approximated with the form in
Equation (\ref{ftest}), \ie the acceleration of pre-existing
and injected CRs along with an exponential cutoff at $p_{\rm max}(t)$.

Comparing the cases in (a) and (b), we see that with the same injection
parameter, the presence of pre-existing CRs results in higher downstream
CR pressure, and that the re-accelerated pre-existing population dominates
over the injected population. 
The presence of pre-existing CRs acts effectively as a
higher injection rate than the thermal leakage alone, leading
to the greatly enhanced CR acceleration efficiency.
For the case with $\epsilon_B=0.3$ in (c), the injection rate is much higher 
than that of the case with $\epsilon_B=0.25$, yet the 
injected population makes a non-negligible contribution only near 
$p_{\rm inj}$.

In Figure 4, we compare the spectrum of re-accelerated CRs from the
steady-state  solutions without injection (left panels) and the CR spectrum
at the shock location from the time-dependent solutions of DSA simulations
at $t/t_o=10$ (right panels),
in order to demonstrate the relative importance of the acceleration of 
the pre-existing and the injected populations.
Different values of $M$ and $s$ are considered, 
but $R=0.05$,  $\delta=0.42$, and $\epsilon_B=0.25$ are fixed. 
As noted before, with the same $R$, the amplitude
$f_{\rm pre}$ of the pre-existing CR spectrum is larger for larger $s$, so
the re-acceleration of pre-existing population is relatively more important.
The figure indicates that for most cases considered, the re-accelerated
pre-existing population dominates over the injected population for the 
considered range of Mach number.
Only for the cases with $s = 4$ and $M \ga 3$,
the freshly injected population makes a noticeable contribution.

Figure 5 shows the downstream CR pressure, $P_{c,2}$, relative to the shock
ram pressure, $\rho_0 u_s^2$, and to the downstream gas thermal pressure,
$P_{g,2}$, as a function of shock Mach number $M$ for different values of
$R$, $s$, and $\delta$.
Again, $\epsilon_B=0.25$ in all the cases.
As shown in the top panels, without pre-existing CRs, both
$P_{c,2}/\rho_0 u_s^2$ and $P_{c,2}/P_{g,2}$ steeply increase with $M$,
because both the injection and acceleration efficiencies depend strongly
on $M$.
For shocks with $M\ga 5$, $P_{c,2}/(\rho_0u_s^2) \ga 0.1$ and the nonlinear
feedback begins to be noticeable.
The feedback reduces the CR injection and saturates the CR acceleration,
so $P_{c,2}$ from DSA simulations becomes smaller than the analytic
estimates in the test-particle limit \citep[see also][]{kr10}.
Also the top panels compare the models with $\delta=0$ and $\delta=0.42$,
demonstrating that the Alfv\'enic drift softens the accelerated spectrum 
and reduces the CR pressure.

In (b) panels, the cases with different upstream CR pressure fractions
are compared:
$P_{c,2}$ increases almost linearly with $R$ at shocks with $M\la 3$ 
in the test-particle regime,
while the CR acceleration begins to show the saturation effect for $M\ga 4$. 
With pre-existing CRs, both $P_{c,2}/\rho_0 u_s^2$ and $P_{c,2}/P_{g,2}$
are substantially larger, compared to the case with $R=0$, 
especially for $M\la 3$, confirming the dominance
of the re-accelerated pre-existing population over the injected population
at weak shocks.

In (c) panels, the cases with different pre-existing slopes are compared;
with softer spectrum (larger $s$), the amplitude $f_{\rm pre}$ is larger
and the CR acceleration is more efficient, as described above with Figure 4.
In (d) panels, the same cases as in (c) panels except $\delta=0$ are shown, 
demonstrating again the effects of Alfv\'enic drift.

These results indicate that at shocks with $M\la 3$ in
ICMs and cluster outskirts, the downstream CR pressure is typically a few
to 10 \% of either the shock ram pressure or the downstream gas thermal
pressure.
Even in the cases where the pre-existing CR population takes up to 10 \% of
the gas thermal pressure in the upstream flow, $P_{c,2}/P_{g,2} \la 0.1$
in the downstream flow. 
This is consistent with the Fermi upper limit \citep{abdo10,ddcb10}.

\section{CR ELECTRONS AND SYNCHROTRON RADIATION}

Since DSA operates on relativistic particles of the same rigidity
($R=pc/Ze$) in the same way, both electrons and protons are expected
to be accelerated at shocks.
However, electrons lose energy, mainly by synchrotron emission and 
Inverse Compton (IC) scattering, and the injection of postshock thermal
electrons is believed to be much less efficient, compared to protons. 

The maximum energy of CR electrons accelerated at shocks can be estimated
by the condition that the momentum gain per cycle by DSA is equal to
the synchrotron/IC loss per cycle, \ie
$\langle \Delta p \rangle_{\rm DSA}+\langle \Delta p \rangle_{\rm rad}$=0
\citep[see][]{wdb84,za07}.
With the assumed Bohm-type diffusion coefficient, the electron spectrum
has a cutoff at
\begin{eqnarray}
p_{\rm cut} & \approx & {m_e^2 c^2 \over \sqrt{4e^3/27}} {u_s \over \sqrt{q}}
\sqrt{B_0 \over B_{0, \rm eff}^2 + B_{2, \rm eff}^2}~~~ ({\rm in~ cgs~ units})
\nonumber \\
& \approx & 340 {{\rm TeV} \over c} \left(u_s \over 10^3{\kms}^{-1}\right)
\sqrt{(B_0/1\muG) \over q\left[(B_{0, \rm eff}/1\muG)^2 +
(B_{2, \rm eff}/1\muG)^2\right]},
\label{pcut}
\end{eqnarray}
where $B_{\rm eff}=(B^2 + B_{\rm CMB}^2)^{1/2}$ with $B_{\rm CMB}=3.24\times
10^{-6}$ G is the effective magnetic field strength for synchrotron and IC
coolings upstream and downstream of shock, and $\delta=0$ was assumed.  
Note that the electron cutoff energy is a time-asymptotic quantity that
depends only on the shock speed and the magnetic field strength,
independent of the shock age.
For a Mach 3 shock and $B_0 = 1\muG$, for example,
the shock jump condition gives 
$\sigma = 3$, $q=4.5$ (with $\delta=0$) and $B_2 = 3\mu$G (assuming
$B \propto \rho$), resulting in the cutoff Lorentz factor,
$\gamma_{\rm e,cut} = p_{\rm cut}/m_e c
\approx 5.6 \times 10^7\ (u_s/1000\kms)$. 

Thus, we may model the downstream electron spectrum as 
\begin{equation}
f_{e,2}(p) \approx K_{e/p}\ f_{p,2}(p)
\exp\left(-{p^2 \over p_{\rm cut}^2}\right),
\label{f2ele}
\end{equation}
where $f_{p,2}(p)$ is the downstream proton spectrum \citep{za07}.
The electron-to-proton number ratio, $K_{e/p}$, is 
not yet constrained precisely by plasma physics
\citep[see, e.g.,][]{reynolds08}.
Although $K_{e/p}\sim 10^{-2}$ is inferred for the Galactic CRs
\citep{schl02}, a much smaller value, $ K_{e/p} \la 10^{-4}$, is preferred
for young supernova remnants \citep{mab09}.
However, $K_{e/p}$ for the pre-existing population in ICMs and cluster
outskirts could be quite different from these estimates.

Next, from the electron spectrum in Equation(\ref{f2ele}), 
we consider the synchrotron emission. 
The averaged rate of synchrotron emission at photon frequency $\nu$ 
from a single
relativistic electron with Lorentz factor $\gamma_e$ can be written as
\begin{equation}
\left<P_{\nu}(\gamma_e)\right> = {4\over3} c \sigma_T \beta^2 U_B \gamma_e^2
\phi_{\nu}(\gamma_e),
\label{pnu}
\end{equation}
where $\beta$ is the particle speed in units of $c$, $\sigma_T$ is the
Thomson cross section, and $U_B$ is the magnetic energy density
\citep[see, e.g.,][]{shu91}.
The frequency distribution function, $\phi_{\nu}(\gamma_e)$, 
which satisfies the normalization $\int \phi_{\nu}(\gamma) d\nu =1$,
peaks at
\begin{equation}
\nu_{\rm peak} \approx \gamma_e^2 \nu_L = 280
\left(B \over 1\ \mu{\rm G}\right)
\left(\gamma_e \over 10^4\right)^2 \ {\rm MHz},
\label{peaknu}
\end{equation}
where $\nu_L=eB/m_e c$ is the Larmor frequency. 
If we approximate that the synchrotron radiation is emitted mostly at
$\nu = \nu_{\rm peak}$ (\ie $\phi_{\nu}(\gamma)$ is replaced by a delta
function centered at $\nu = \nu_{\rm peak}$), the synchrotron volume
emissivity from the CR electron number density,
$n_e(\gamma_e) d\gamma_e = f_e(p)p^2dp$, becomes
\begin{equation}
J({\nu})\approx {2\over3} c \sigma_T \beta^2 U_B
{ \gamma_e \over \nu_L} n_e(\gamma_e),
\label{synemi}
\end{equation}
with $\gamma_e$ corresponding to the given $\nu_{\rm peak} = \nu$
in Equation (\ref{peaknu}).
So the ratio of the downstream to upstream synchrotron emissivity at 
a given frequency $\nu$ can be written as
\begin{equation}
{J_2({\nu}) \over J_0({\nu})}  \approx {B_2 \over B_0}
{\gamma_{e,2}^3 f_{e,2}(\gamma_{e,2}) \over \gamma_{e,0}^3
f_{e,0}(\gamma_{e,0})},
\label{emirat}
\end{equation}
where $\gamma_{e,0}$ and $\gamma_{e,2}$ are the Lorenz factor
that corresponds to the given $\nu_{\rm peak}=\nu$ in Equation
(\ref{peaknu}) for upstream field $B_0$ and downstream field $B_2$,
respectively. 

For power-law spectra, the ratio $J_2({\nu})/J_0({\nu})$ can be written
in a more intuitive form.
If the ratio $K_{e/p}$ of the pre-existing population is comparable to 
or greater than that of the injected population, 
pre-existing electrons are more important than 
injected electrons at weak shocks of $M \la 3$,
as pointed out in the previous section. 
Then, the downstream electron spectrum $f_{e,2}$ can be approximated 
by the distribution function
in Equation (\ref{f2p}) with a Gaussian cutoff, $\exp(-p^2/p_{\rm cut}^2)$.
Again adopting $f_{e,0}(\gamma_e) \propto \gamma_e^{-s}$ for pre-existing 
CR electrons, the downstream spectrum is 
$f_{e,2}(\gamma_e) \propto \gamma_e^{-r}$, (unless $q=s$) for
$\gamma_e < \gamma_{\rm e, cut} \equiv p_{\rm cut}/m_e c$.
Then, the ratio of the downstream to upstream synchrotron emissivity at
$\nu$ becomes
\begin{eqnarray}
{J_2({\nu}) \over J_0({\nu})} & \approx & 
\left(B_{2,\mu{\rm G}}^{(r-1)/2} \over B_{0,\mu{\rm G}}^{(s-1)/2}\right)
\left[{f_{e,2}(\gamma_e) \over f_{e,0}(\gamma_e)}\right]_{\gamma_e=10^4}
\left(\nu \over 280\ {\rm MHz} \right)^{(s-r)/2} \nonumber \\
& \approx & \sigma^{w(r-1)/2} B_{0,\mu{\rm G}}^{-(s-r)/2}
\left[{f_{e,2}(\gamma_e) \over f_{e,0}(\gamma_e)}\right]_{\gamma_e=10^4}
\left(\nu \over 280\ {\rm MHz}\right)^{(s-r)/2},
\label{emirat2}
\end{eqnarray}
where $B_{0,\mu{\rm G}}$ and $B_{2,\mu{\rm G}}$ are the upstream and
downstream magnetic field strengths in units of $\mu$G.
In the second step, we assumed that $B_2/B_0 = (\rho_2/\rho_0)^w = \sigma^w$,
where $w=1$ corresponds to $B \propto \rho$ implied by the diffusion model 
in Equation (\ref{diffcoef}), 

Figure 6 shows $f_{e,0}(\gamma_e)/f_{e,2}(\gamma_e)$ at $\gamma_e=10^4$, and
$(J_2/J_1)_{280} \equiv J_2({\nu}) / J_0({\nu})$ at $\nu = 280$ MHz  
for $B_0 = 1 \mu$G and $w = 1$ for the cases considered in Figure 5.
Here, we assume that $K_{e/p}$ is the same for both the pre-existing 
and injected populations. 
Since the electron cutoff momentum is $\gamma_{\rm cut}\sim 10^8$ 
for the shock parameters considered here, the choice of $\gamma_e=10^4$
and $\nu = 280$ MHz (see Equation(\ref{peaknu})) as the representative
values should be safe.
As shown in Figure 5, for $M \la 3$, 
the downstream CR proton pressure can absorb typically only a few to 10\%
of the shock ram pressure even for $R=0.05$. 
Yet, the acceleration of CR electrons can result in a substantial
enhancement in synchrotron radiation across the shock.
Our estimation indicates that the enhancement factor, $(J_2/J_1)_{280}$, 
can be up to several at shocks with $M \sim 1.5$,
up to several 10s for $M \sim 2$, and up to several 100s for $M \sim 3$.
This is partly due to the large enhancement of the electron population
across the shock, $f_{e,2}/f_{e,0}$, 
which is typically an order of magnitude smaller than the ratio
$(J_2/J_0)_{280}$.
Additional enhancement comes from the amplification of magnetic fields
across the shock, $B_2/B_0$.

We note that for the compression of a uniform magnetic field,
$B \propto \rho^{2/3}$, that is, $w=2/3$.
With this scaling, $(J_2/J_0)_{280}$ should be a bit smaller than 
that in Figure 6.
However, it is also quite plausible that the downstream magnetic field
is stronger than that expected for simple compression.
It has been suggested that at shocks, especially at strong shocks, the
downstream magnetic field is amplified by plasma instabilities
\citep[see, e.g.,][]{lucek00,bell04}, although the existence of such
instabilities has not been fully explored for weak shocks.
Moreover, the magnetic field can be further amplified by the turbulence
that is induced through cascade of the vorticity generated behind shocks
\citep{giacal07,ryuetal08}.
In such cases, the ratio $(J_2/ J_0)_{280}$ could be larger than that
in Figure 6.
In that sense, our estimate for the synchrotron enhancement factor
may be considered as conservative one.
We also note that with $s \geq r$ in Equation (\ref{emirat2}),
$J_2({\nu}) / J_0({\nu})$ is larger at higher frequencies, but smaller
with larger $B_0$.

The above enhancement in synchrotron emission across the shock can be
compared to the enhancement in Bremsstrahlung X-ray.
The Bremsstrahlung X-ray emissivity is given as $J_X \propto \rho^2\sqrt{T}$,
so the ratio of the downstream to upstream emissivity can be written as
\begin{eqnarray}
{J_{X,2} \over J_{X,0}} & = & \sigma^2 \sqrt{T_2 \over T_0} \nonumber \\
& = & \left(4M^2 \over M^2+3\right)^{3/2} \left(5M^2-1 \over 4 \right)^{1/2},
\label{xrayrat}
\end{eqnarray}
in the limit where the CR pressure does not modify the shock structure.
The enhancement in Bremsstrahlung X-ray emission, 
$J_{X,2} / J_{X,0}$,
is 3.6, 7.5, and 17 for $M = 1.5$, 2, and 3, respectively.
These values are substantially smaller than $(J_2/J_0)_{280}$ shown 
in Figure 6. 
This implies that shocks in ICMs and cluster outskirts may appear as radio
relics, but not be detected in X-ray, for instance, as in the case of 
CIZA J2242.8+5301 \citep{wrbh10}.

Since the synchrotron/IC cooling time scales as 
\begin{equation}
t_{\rm rad}= {2.45\times 10^{13} \yrs\ \over  \gamma_e}
\left(B_{\rm eff,2} \over 1\muG\right)^{-2}
\label{radtime}
\end{equation}
\citep{wdb84}, behind the shock the width of the distribution of CR
electrons with $\gamma_e$ becomes
$d \approx u_2 t_{\rm rad}(\gamma_e) \propto \gamma_e^{-1}$.
For instance, electrons radiating synchrotron at $\nu \sim 1$ GHz have
mostly the Lorentz factor of $\gamma_e \approx 10^4$ in the magnetic field
of $B_2 \sim$ a few $\mu$G.
So the width of the synchrotron emitting region behind the shock
is $d \approx u_2 t_{\rm rad}(\gamma_e=10^4)
\sim 100\ {\rm kpc}\ (u_2/10^3\kms)$
as long as the shock age $t> t_{\rm rad}(\gamma_e=10^4)\sim 10^8 \yrs$.
This is indeed of order the width of bright radio relics such as
CIZA J2242.8+5301 \citep{wrbh10}.

Moreover, from the fact that $d\propto \gamma_e^{-1}$, 
we can identify another feature in the integrated synchrotron spectrum.
The volume integrated electron spectrum,
$F_{e,2}(\gamma_e) \propto f_{e,2}(\gamma_e) \cdot d \propto
\gamma_e^{-(r+1)}$, 
steepens by one power of $\gamma_e$ above the break Lorentz factor,
$\gamma_{\rm e,br} \approx 2.45 \times 10^5 \left({{10^8 \yrs}
/ t}\right) \left({B_{\rm eff,2} / {1\muG}}\right)^{-2}$,
where $t$ is the shock age.
Note that the break Lorentz factor is basically derived 
from the condition, $t=t_{\rm rad}$ in Equation (\ref{radtime}) and 
so independent of the shock speed.
Hence, if $f_{e,2}(\gamma_e) \propto \gamma_e^{-r}$,
in observations of unresolved sources, 
the integrated synchrotron emission, $S_{\nu}\propto \nu^{-\alpha}$, 
has the spectral slope $\alpha=(r-3)/2$ for $\nu < \nu_{\rm br}$
and $\alpha=(r-2)/2$  for $\nu_{\rm br} \la \nu \la \nu_{\rm cut}$.
Here, the two characteristic frequencies, 
$\nu_{\rm br}$ and $\nu_{\rm cut}$, 
correspond to the peak frequency in Equation (\ref{peaknu}) for 
$\gamma_{\rm e,br}$ and $\gamma_{\rm e,cut}$, respectively.
So the spectral slope of the integrated spectrum just below the cutoff 
frequency is steeper by 0.5 than that of the resolved spectrum. 
 
\section{SUMMARY}

Cosmological shocks are expected to be present 
in the large-scale structure of the universe.
They form typically with Mach number up to $10^3$ and speed up to
a few 1000 km s$^{-1}$ at the present universe.
Shocks in ICMs and cluster outskirts with relatively high X-ray 
luminosity, in particular,
have the best chance to be detected, so they have started
to be observed as X-ray shocks and radio relics (see Introduction for
references).
Those shocks are mostly weak with small Mach number of 
$M\la 3$, because they form in the hot gas of $kT \ga$ keV.

In this paper, we have studied DSA at weak cosmological shocks.
Since the test-particle solutions could provide a simple yet reasonable
description for weak shocks, we first suggested analytic solutions which
describe the {\it time-dependent} DSA in the test-particle regime,
including both the {\it pre-existing} and injected CR populations.
We adopted a thermal leakage injection model to emulate the acceleration
of suprathermal particles into the CR population, along with a simple
transport model in which Alfv\'en waves self-excited by the CR streaming
instability drift relative to the bulk plasma upstream of the gas subshock.

We then performed kinetic DSA simulations and compared 
the analytic and numerical solutions for wide ranges of model parameters
relevant for shocks in ICMs and cluster outskirts: 
the shock Mach number $M = 1.5 - 5$, the slope of the 
pre-existing CR spectrum $s=4-5$, the ratio of the upstream CR to gas
pressure $R=0.01-0.1$, the injection parameter $\epsilon_B=0.25-0.3$,
and the Alfv\'enic speed parameter $\delta=0-0.42$.
The upstream gas was assumed to be fully ionized with $T_0=10^7$ K.

The main results can be summarized as follows:

1) For weak shocks with $M \la 3$, 
the test-particle solutions given in Equation (\ref{ftest}) should
provide a good approximation for the time-dependent CR spectrum at
the shock location.
We note that the test-particle slope, $q$, in Equation (\ref{qtp})
and the maximum momentum, $p_{\rm max}(t)$, in Equation (\ref{pmax})
may include the Alfv\'enic drift effect.

2) For the injection parameter considered here, $\epsilon_B=0.25-0.3$, the
injection fraction is rather low, typically $\xi \sim 5\times10^{-5}$ to
$10^{-3}$ for $M \la 3$.
The pre-existing CR population provides more particles for DSA than
the freshly injected population.
Hence, the pre-existing population dominates over the injected population.
If there exist no CRs upstream ($R=0$), the downstream CR pressure absorbs
typically much less than $\sim 1$ \% of the shock ram pressure for $M \la 3$.
With pre-existing CRs that accounts for 5 \% of the gas 
thermal pressure in the upstream flow, 
the CR acceleration efficiency increases to a few to 10 \% for those
weak shocks.

3) For the pre-exisiting population, 
the enhancement of the distribution function across the shock,
$f_2(p)/f_1(p)$, at a given momentum is substantially larger than that
expected from the simple adiabatic compression.
Hence, with amplified magnetic fields downstream,
the re-acceleration of pre-existing CR electrons can result in a
substantial synchrotron radiation behind the shock.
We estimated that the enhancement in synchrotron radiation across the
shock, $J_2({\nu})/J_0({\nu})$, is about
a few to several for $M \sim 1.5$, while it could reach to $10^2-10^3$
for $M\sim 3$, depending on the detail model parameters.
This is substantially larger than the enhancement in X-ray emission.

4) Unlike protons, relativistic electrons lose energy by synchrotron
emission and IC scattering behind the shock, resulting in a finite width
of synchrotron emitting region.
In ICMs and cluster outskirts with $\mu$G fields, 
the radio synchrotron emission at $\nu\sim 1$GHz originate mostly from
the relativistic electrons with $\gamma_e\sim 10^4$, which cool in a time
scale of $t_{\rm rad}\sim 10^8$ yrs. 
So the width of the $\sim 1$ GHz synchrotron emitting region is 
$d\approx u_2 t_{\rm rad} \sim 100~{\rm kpc}\ (u_s/1000\kms)$
for a shock of age $t>t_{\rm rad}$.

Finally, although the CRASH numerical code and our thermal leakage model
are developed for quasi-parallel shocks, the main conclusions in this paper
should be valid for quasi-perpendicular shocks as well.
It is recognized that the injection may be less efficient and
the self-excited waves are absent at perpendicular shocks.
However, both of these problems are alleviated in the presence of
pre-existing CRs and turbulence \citep{giacal05,zank06}. 
So the diffusion approximation should be valid and the re-acceleration
of pre-existing CRs are similar at both kinds of shocks.
Then, we expect our results can be applied to, for instance,
CIZA J2242.8+5301, the radio relic whose magnetic field direction
inferred from the polarization observation is perpendicular to
the shock normal. 

\acknowledgements
The authors would like to thank T. W. Jones and J. Cho for discussions.
HK was supported by Basic Science Research Program through
the National Research Foundation of Korea (NRF) funded by the Ministry
of Education, Science and Technology (2010-0016425).
DR was supported by the National Research Foundation of Korea
through grant 2007-0093860.

\clearpage

\begin{figure}
\vspace{-5cm}
\hskip -1cm
\includegraphics[scale=0.85]{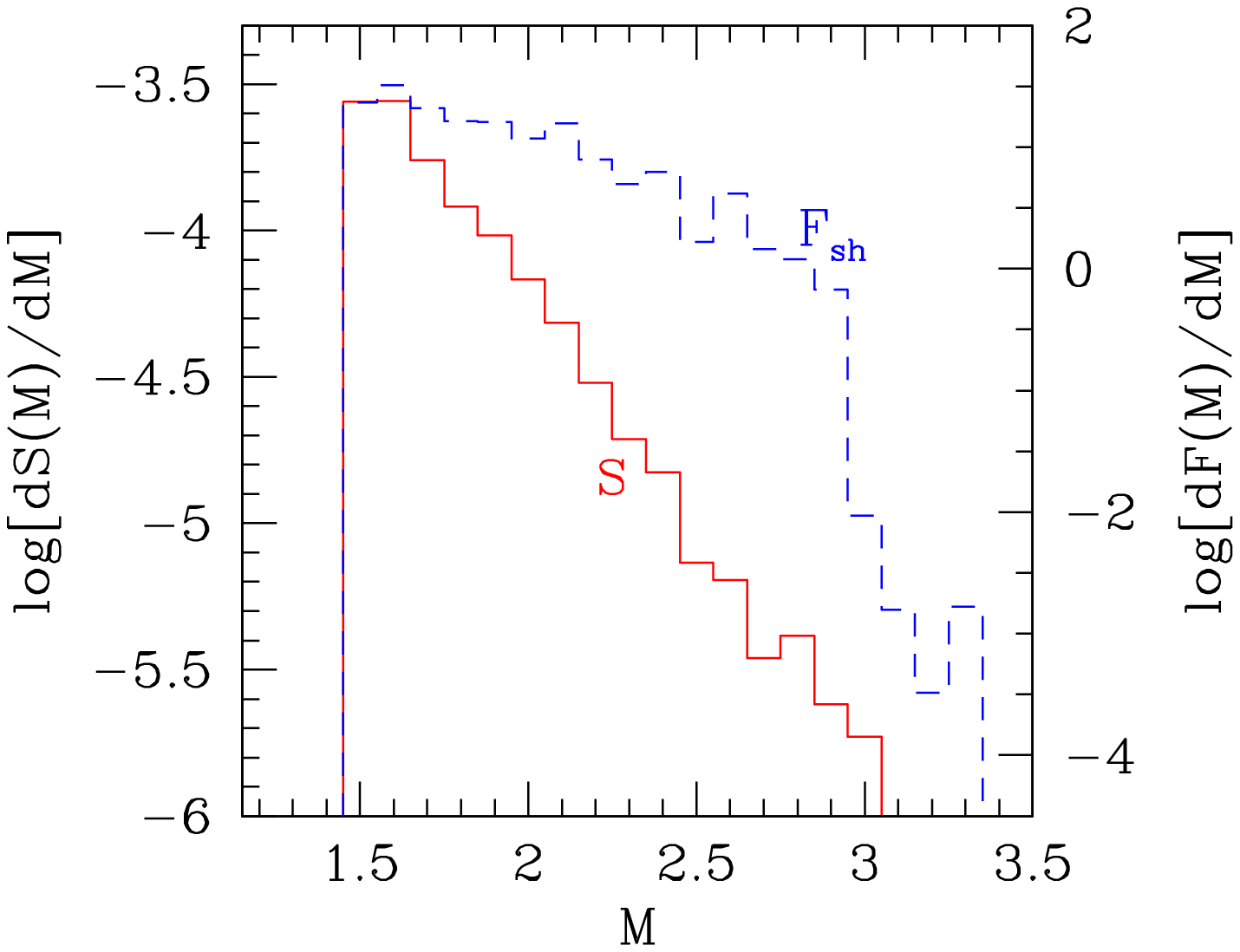}
\vspace{-3cm}
\caption{Surface area of shocks in ICMs and cluster outskirts, $S$ (red
solid line), and kinetic energy flux passed through surfaces of the
shocks, $F_{sh}$ (blue dashed line), as a function of Mach number $M$
at $z=0$.
Only shocks with the preshock gas temperature of $T_1 \ge 10^7$ K are
considered.}
\end{figure}

\clearpage

\begin{figure}
\vspace{-5cm}
\begin{center}
\includegraphics[scale=0.85]{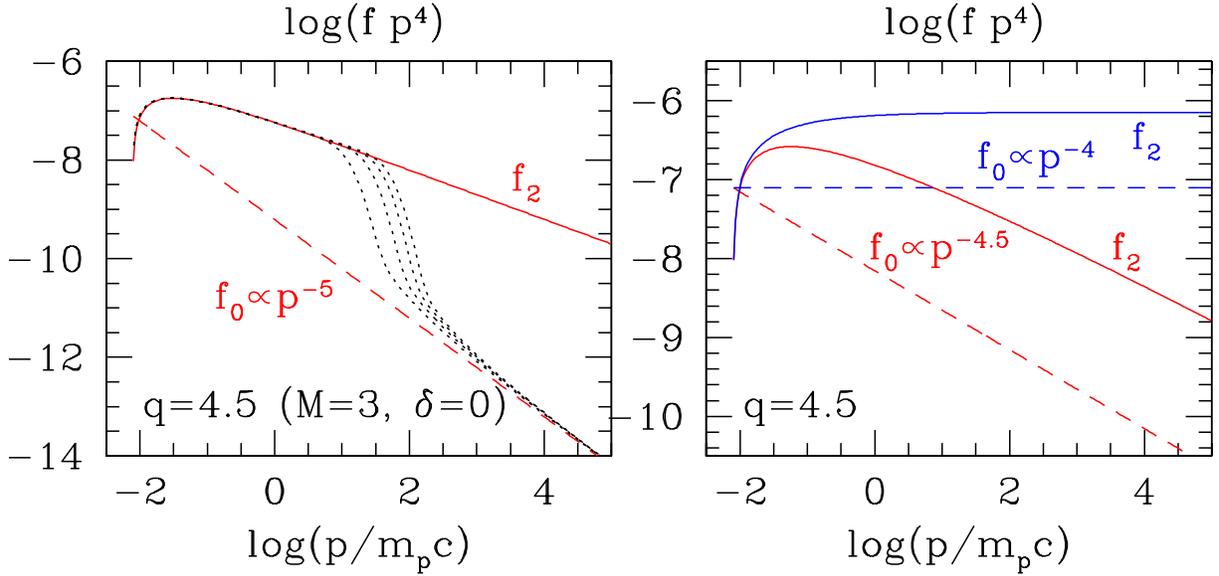}
\end{center}
\vspace{-5cm}
\caption{Steady-state solution for the downstream CR spectrum, 
$f_2^{\rm reac}(p)$, given in Equation (\ref{f2p}) (sold lines),
accelerated from an upstream CR spectrum, $f_0(p) \propto p^{-s}$
(dashed lines). 
A shock with Mach number $M = 3$ is considered, so the test-particle
slope is $q=4.5$ (with $\delta \equiv v_A/c_s = 0$).
The CR injection is ignored and the distribution function $f(p) p^4$
is plotted.
Left: The case with the slope $s=5$. 
The dotted lines show the time-dependent solution at the shock location,
$f_s(p)$, from the corresponding DSA simulation. 
Right: The case with slope $s=4$ and 4.5.}
\end{figure}

\clearpage

\begin{figure}
\vspace{-2cm}
\begin{center}
\includegraphics[scale=0.85]{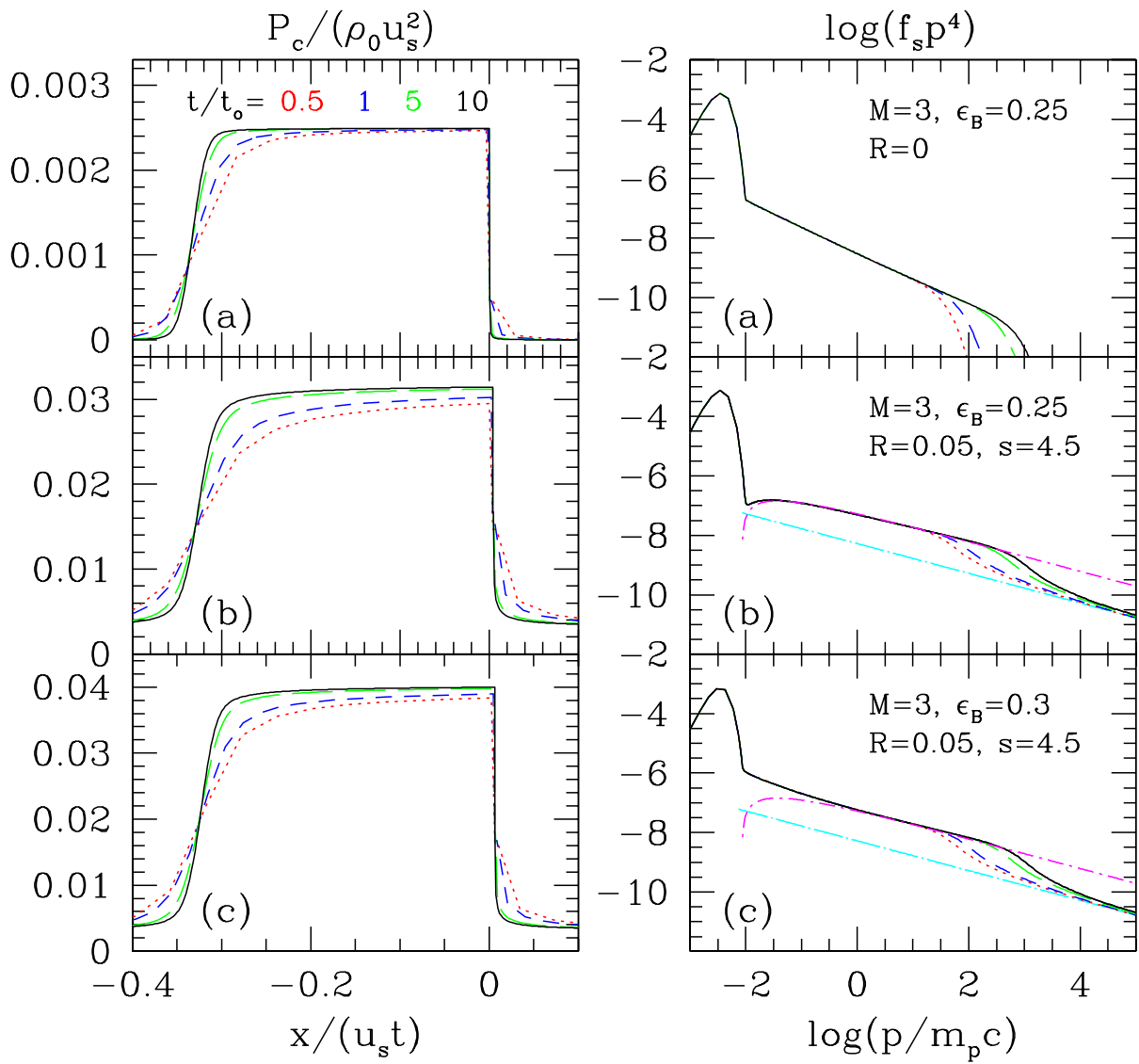}
\end{center}
\vspace{-1.5cm}
\caption{Time-dependent solution from DSA simulations of a Mach 3
shock.
The CR pressure profile as a function of the similarity variable
$x/(u_st)$ (right panels) and the CR distribution at the shock location,
$f_s(p)$, (left panels) at $t/t_o=0.5$ (red dotted lines), 1 (blue
dashed lines), 5 (green long dashed line), and 10 (black solid lines)
for three cases.
The top panels show the case with the injection parameter $\epsilon_B = 0.25$
and without pre-existing CRs.
The middle panels show the case with $\epsilon_B=0.25$ and with
pre-existing CRs: the ratio of the upstream, pre-existing CR to gas pressure
is $R \equiv P_{c,0}/P_{g,0} = 0.05$ and the spectral slope of the
pre-existing CRs is $s=4.5$.
The bottom panels show the case with the same pre-existing CRs, but
with a higher injection rate, $\epsilon_B=0.3$.
All cases shown have $\delta \equiv v_A/c_s =0.42$.  
In the left panels, the CR pressure is displayed in different vertical
scales for clarity.
In the right panels, for the cases (b) and (c), the (magenta) dot-dashed
lines show the steady-state solution of the re-accelerated CRs,
$f_2^{\rm reac}(p)$, in Equation (\ref{f2p}), while the (cyan) dot-long
dashed lines show the pre-existing CR spectrum, $f_0(p)$, for comparison.}
\end{figure}

\clearpage

\begin{figure}
\vspace{-2cm}
\begin{center}
\includegraphics[scale=0.85]{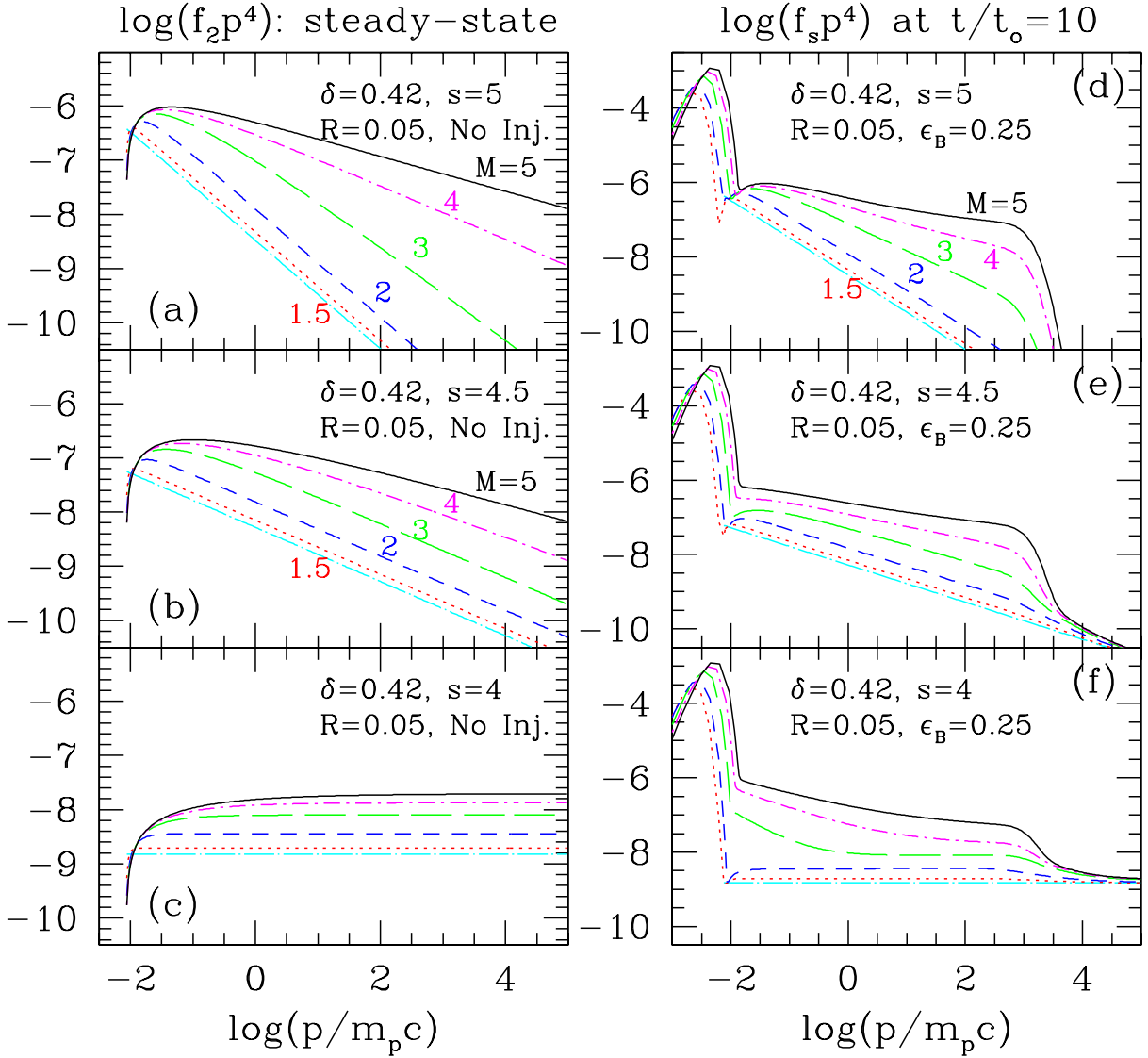}
\end{center}
\vspace{-1.5cm}
\caption{Left: Steady-state solution of the re-accelerated CR spectrum
(without injected population), $f_2^{\rm reac}(p)$, in Equation (\ref{f2p}).
Right: Time-dependent solution of the CR distribution at the shock location, 
$f_s(p)$, at $t/t_o=10$ from DSA simulations with the injection parameter 
$\epsilon_B = 0.25$.
Three different spectral slopes of pre-existing CRs $s=$ 4, 4.5, 5
are considered.
In all cases, $R = P_{c,0}/P_{g,0} = 0.05$ and $\delta = v_A/c_s = 0.42$. 
Results are shown for shocks with Mach number $M=1.5$ (red dotted lines), 2
(blue dashed lines), 3 (green long dashed lines), 4 (magenta dot-dashed
line), 5 (black solid lines).
The (cyan) dot-long dashed lines plot the pre-existing CRs, $f_0(p)$.}
\end{figure}

\clearpage

\begin{figure}
\vspace{-1cm}
\begin{center}
\includegraphics[scale=0.85]{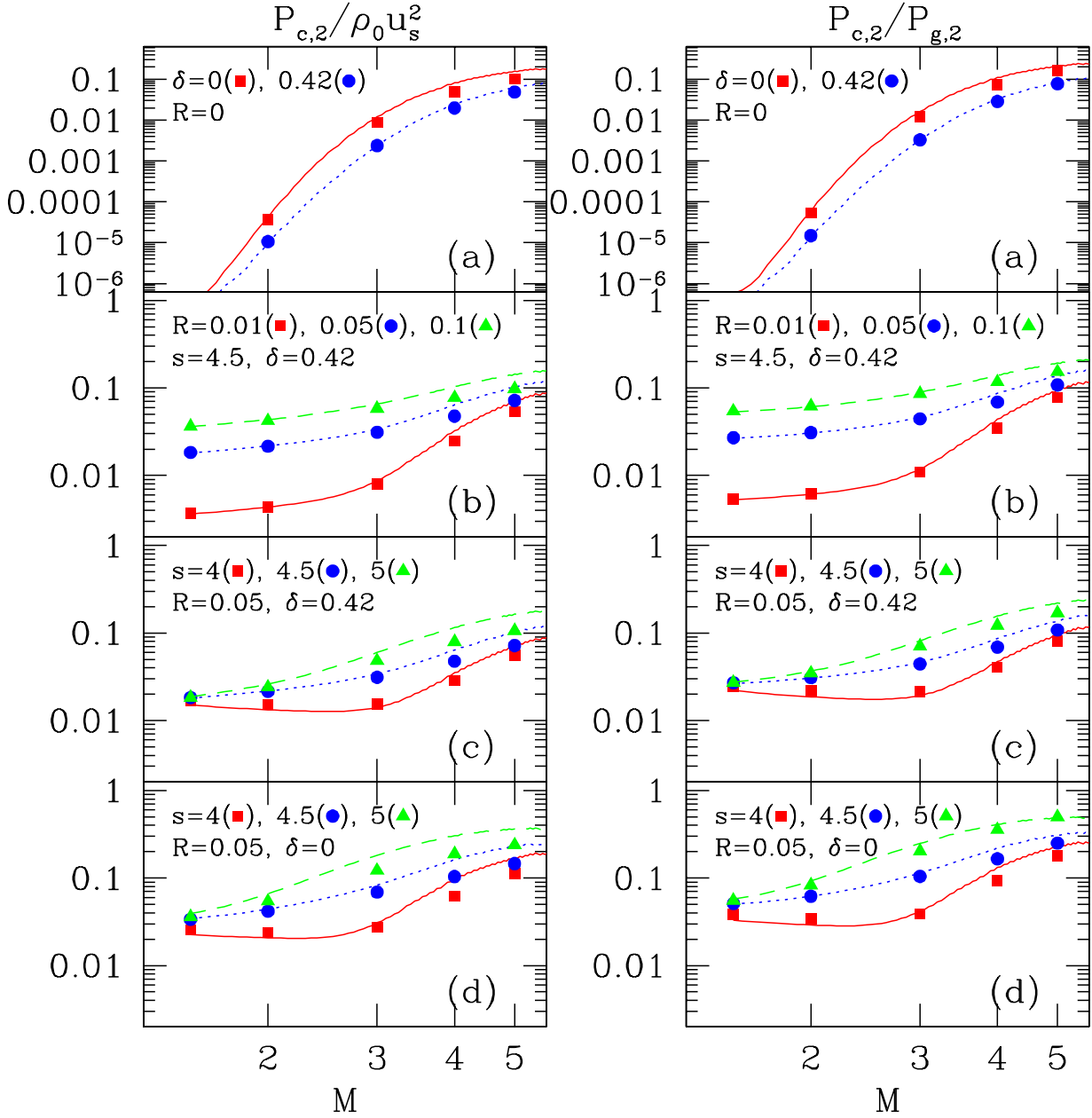}
\end{center}
\vspace{-1cm}
\caption{Ratios of the downstream CR pressure to the shock ram pressure 
(left panels) and to the downstream gas thermal pressure (right panels)
for different model parameters. 
Lines show the ratios estimated from the analytic formula in Equation
(\ref{ftest}), while symbols show the time-asymptotic values from the
corresponding DSA simulations at $t/t_o=10$.  
The top panes show the cases without pre-existing CRs for two different
$\delta \equiv v_A/c_s =0$ and $0.42$.
The second panels from top show the cases with pre-existing CRs of different
$R \equiv P_{c,0}/P_{g,0} = 0.01$, 0.05 and 0.1.
The spectral slope of pre-existing CRs is $s=4.5$, and $\delta=0.42$ is
adopted.
The third panels from top show the cases with pre-existing CRs of
different $s=4$, $4.5$ and $s=5$.
Other parameters are $R = 0.05$ and $\delta = 0.42$.
The bottom panels show the same cases as the third panels except
$\delta = 0$.
In all cases, $\epsilon_B=0.25$ is used.}
\end{figure}

\clearpage

\begin{figure}
\vspace{-2cm}
\begin{center}
\includegraphics[scale=0.85]{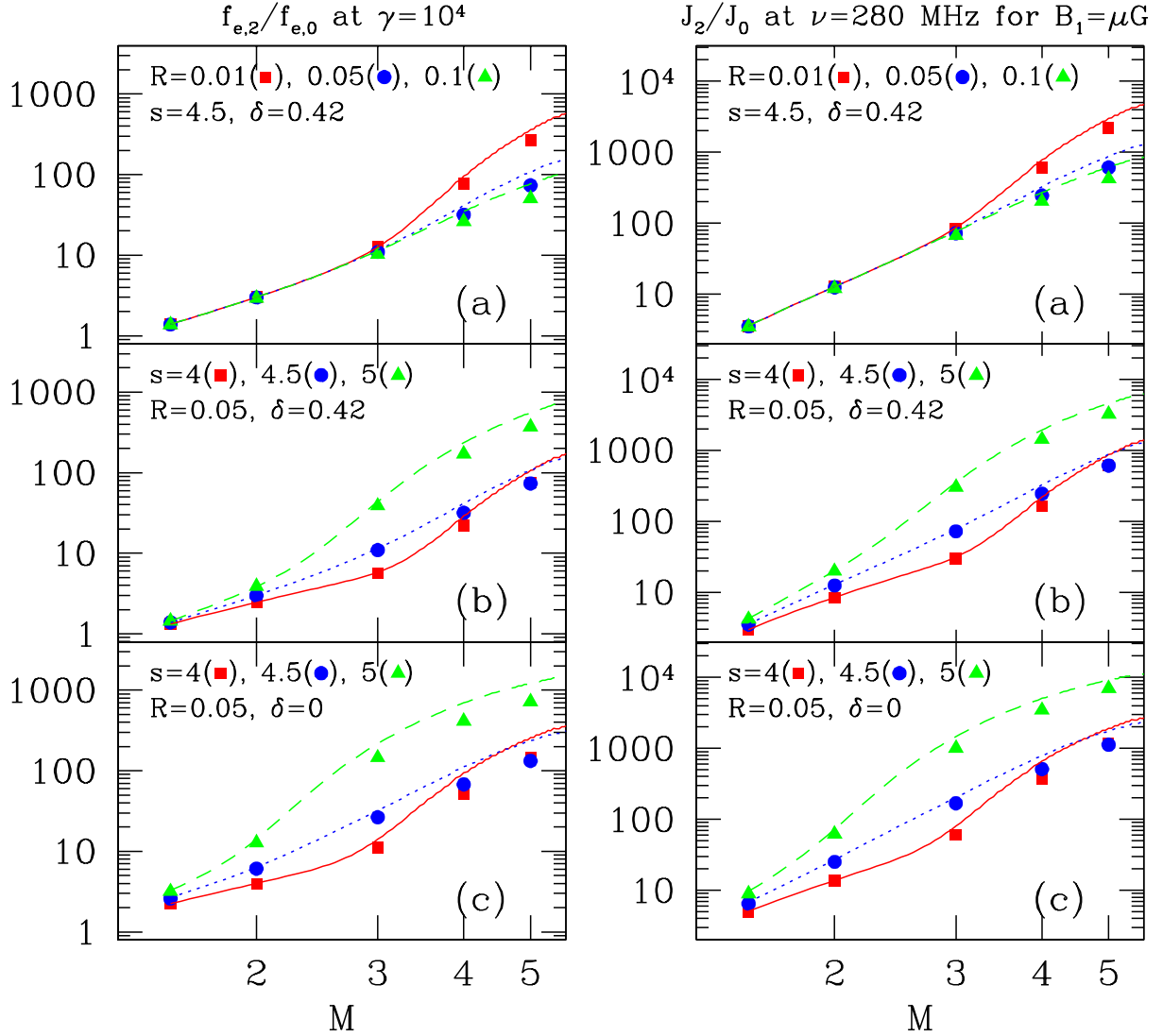}
\end{center}
\vspace{-1.5cm}
\caption{Ratios of the downstream to upstream CR electrons, 
$f_{e,2}(p)/f_{e,0}(p)$ at $p=10^4\ m_e c$ (left panels)
and the downstream to upstream synchrotron emissivity,
$J_2(\nu)/J_0(\nu)$ at $\nu=280$ MHz (right panels).
The synchrotron emissivity was calculated by Equation (\ref{emirat})
for the upstream and downstream magnetic field, $B_0 = 1\muG$, 
and $B_2 = (\rho_2/\rho_0) B_0$, respectively.
The same cases as in Figure 5 except the one without pre-existing CRs 
are shown here.
See the caption for Figure 5 for different line and symbol types.}
\end{figure}


\begin{thebibliography}{}

\bibitem[Abdo et al.(2010)]{abdo10}
Abdo, A. A. et al.
2010, Science, 327, 1103

\bibitem[Ackermann et al.(2010)]{acke10}
Ackermann, M. et al.
2010, \apjl, 717, L71

\bibitem[Amato \& Blasi(2006)]{ab06}
Amato, E., \& Blasi, P.
2006, \mnras, 371, 1251

\bibitem[Bagchi et al.(2006)]{bdnp06}
Bagchi, J., Durret, F., Neto, G. B. L., \& Paul, S.
2006, Science, 314, 791

\bibitem[Bell(1978)]{bell78}
Bell, A. R.
1978, \mnras, 182, 147

\bibitem[Bell(2004)]{bell04}
Bell, A.R.,
2004, \mnras, 353, 550

% \bibitem[Blumenthal \& Gould(1970)]{bg70}
% Blumenthal, G. R., \& Gould, R. J.
% 1970, Rev. Mod. Phys., 42, 237

% \bibitem[Brunetti \& Blasi(2005)]{brunetti05}
% Brunetti G., \& Blasi P.
% 2005, \mnras, 363, 1173

\bibitem[Brunetti \& Lazarian(2007)]{bl07}
Brunetti G., \&  Lazarian, A.
2007, \mnras, 378, 245

\bibitem[Caprioli et al.(2009)]{capri09} 
Caprioli, D., Blasi, P., \&  Amato, E.
2009, \mnras, 396, 2065

\bibitem[Carilli \& Taylor(2002)]{ct02}
Carilli, C. L., \& Taylor, G. B.
2002, \araa, 40, 319

\bibitem[Cassano \& Brunetti(2005)]{cb05} 
Cassano, R., \& Brunetti, G.
2005, \mnras, 357, 1313

% \bibitem[Cassano et al.(2007)]{cassano07} 
% Cassano, R., Brunetti, G., Setti, G., Govoni F., \& Dolag K.
% 2007, \mnras, 378, 1565

\bibitem[Cen \& Ostriker(1999)]{co99}
Cen, R., \& Ostriker, J. P.
1999, \apj, 514, 1

\bibitem[Cen \& Ostriker(2006)]{co06}
Cen, R., \& Ostriker, J. P.
2006, \apj, 650, 560

\bibitem[Chandran(2005)]{chan05} 
Chandran, B. D. G.
2005, \prl, 95, 265004

\bibitem[Donnert et al.(2010)]{ddcb10} 
Donnert, J., Dolag, K., Cassano, R., \& Brunetti, G.
2010, \mnras, 407, 1565

\bibitem[Drury(1983)]{dru83} 
Drury, L. O'C.
1983, Rept. Prog. Phys., 46, 973

% \bibitem[Feretti(2005)]{feretti05}
% Feretti, L.
% 2005, Adv. Space Res., 36, 729

% \bibitem[Feretti et al.(2004)]{feretti04}
% Feretti, L., Brunetti, G., Giovannini, G., Kassim, N., Orr\'u, E.,
% \& Setti, G.
% 2004, J. Korean Astron. Soc., 37, 315

\bibitem[Finoguenov et al.(2010)]{fsnw10}
Finoguenov, A., Sarazin, C. L., Nakazawa, K. Wik, D. R., \& Clarke, T. E.
2010, \apj, 715, 1143

\bibitem[Giacalone(2005)]{giacal05}
Giacalone, J.,
2005, \apj, 628, L37

\bibitem[Giacalone \& Jokipii(2007)]{giacal07}
Giacalone, J., \& Jokipii, J. R. 
2007, \apjl, 663, L41

\bibitem[Govoni \& Feretti(2004)]{gf04}
Govoni, F., \& Feretti, L.
2004, Int. J. Mod. Phys. D, 13, 1549

\bibitem[Hoeft et al.(2008)]{hoeft08} 
Hoeft, M., Bruggen, M., Yepes, G., Gottlober, S., \& Schwope, A.,
2008, \mnras, 391, 1511

\bibitem[Kang et al.(2002)]{kjg02}
Kang, H., Jones, T. W., \& Gieseler, U. D. J.
2002, \apj, 579, 337

\bibitem[Kang \& Jones(2007)]{kj07}
Kang, H., \& Jones, T. W.
2007, Astropart. Phys., 28, 232

\bibitem[Kang \& Ryu(2010)]{kr10}
Kang, H., \& Ryu, D.
2010, \apj, 721, 886

\bibitem[Kang et al.(2007)]{kangetal07}
Kang, H., Ryu, D., Cen, R., \& Ostriker, J. P.
2007, \apj, 669, 729

\bibitem[Kang et al.(2005)]{krcs05}
Kang, H., Ryu, D., Cen, R., \& Song, D.
2005, \apj, 620, 21

\bibitem[Kang et al.(2009)]{krj09}
Kang, H., Ryu, D., \& Jones, T. W. 
2009, \apj, 695, 1273

% \bibitem[Lang(1999)]{lang99}
% Lang, K. R.
% 1999, Astrophysical Formulae Volume 1: Radiation, Gas Processes and
% High Energy Astrophysics (Berlin: Springer), p. 37

\bibitem[Lucek \& Bell(2000)]{lucek00}
Lucek, S. G., \& Bell, A. R.
2000, \mnras, 314, 65

% \bibitem[Longair(1994)]{long94}
% Longair, M. S.
% 1994, High Energy Astrophysics Volume 1: Particles, Photons and
% Their Detection (Cambridge: Cambridge Univ. Press)

% \bibitem[Malkov(1998)]{mal98} 
% Malkov M. A.
% 1998, \pre, 58, 4911

\bibitem[Malkov \& V\"olk(1998)]{mv98}
Malkov, M. A., \& V\"olk, H. J.
1998, Adv. Space Res., 21, 551
%Values for epsilon parameter

\bibitem[Malkov \& Drury(2001)]{maldru01} 
Malkov M. A., \& Drury, L. O'C. 
2001, Rep. Prog. Phys., 64, 429
%CR general review with injection

\bibitem[Markevitch et al.(2002)]{markev02}
Markevitch, M., Gonzalez, A. H., David, L., Vikhlinin, A., Murray, S.,
Forman, W., Jones, C., \& Tucker, W.
2002, \apjl, 567, L27

\bibitem[Markevitch et al.(2005)]{markev05}
Markevitch, M., Govoni, F., Brunetti, G., \& Jerius, D.
2005, \apj, 627, 733

\bibitem[Markevitch \& Vikhlinin(2007)]{markev07}
Markevitch, M., \& Vikhlinin, A.
2007, Phys. Rep., 443, 1

% \bibitem[Million \& Allen(2009)]{million09}
% Million, E. T., \& Allen, S. W.
% 2009, \mnras, 399, 1307

\bibitem[Miniati et al.(2000)]{miniati00}
Miniati, F., Ryu, D., Kang, H., Jones, T. W., Cen, R.,
\& Ostriker, J. P.  
2000, \apj, 542, 608

\bibitem[Morlino et al.(2009)]{mab09}
Morlino G., Amato E., \& Blasi P.
2009, \mnras, 392, 240

\bibitem[Pfrommer et al.(2006)]{psej06}
Pfrommer, C., Springel, V., En{\ss}lin, T. A., \& Jubelgas, M.
2006, \mnras, 367, 113

% \bibitem[Pfrommer et al.(2007)]{pfrommer07}
% Pfrommer, C., En{\ss}lin, T. A., Springel, V., Jubelgas, M.,
% \& Dolag, K.
% 2007, \mnras, 378, 385

% \bibitem[Ptuskin et al.(2010)]{pzs10} 
% Ptuskin, V. S., Zirakashvili, V. N., \& Seo, E.S.
% 2010, \apj, 718, 31

\bibitem[Parizot et al.(2006)]{parizot06}
Parizot, E., Marcowith, A., Ballet, J., \& Gallant, Y. A. 
2006, \aap, 453, 387

\bibitem[Reynolds(2008)]{reynolds08}
Reynolds, S. P.
2008, \araa, 46, 89

\bibitem[Ryu et al.(2008)]{ryuetal08}
Ryu, D., Kang, H., Cho, J., \&  Das, S.
2007, Science, 320, 909

\bibitem[Ryu et al.(2003)]{ryuetal03}
Ryu, D., Kang, H., Hallman, E., \& Jones, T. W.
2003, \apj, 593, 599

\bibitem[Ryu et al.(1993)]{rokc93}
Ryu, D., Ostriker, J. P., Kang, H., \& Cen, R.
1993, \apj, 414, 1

\bibitem[Schlickeiser(2002)]{schl02}
Schlickeiser R.
2002, Cosmic Ray Astrophysics (Berlin: Springer)

\bibitem[Shu(1991)]{shu91}
Shu, F. H.
1991, The Physics of Astrophysics Volume 1: Radiation
(Mill Valley: University Science Books)

\bibitem[Skilling(1975)]{skill75} 
Skilling, J. 
1975, \mnras, 172, 557

\bibitem[Skillman et al.(2008)]{skillman08}
Skillman, S. W., O'Shea, B. W., Hallman, E. J., Burns, J. O.,
\& Norman, M. L.
2008, \apj, 689, 1063

\bibitem[Vazza et al.(2009)]{vazza09}
Vazza, F., Brunetti, G., \& Gheller, C.
2009, \mnras, 395, 1333

\bibitem[van Weeren et al.(2010)]{wrbh10}
van Weeren, R., R\"ottgering, H. J. A., Br\"uggen, M., \& Hoeft, M.
2010, Science, 330, 347

\bibitem[Vladimirov et al.(2006)]{veb06}
Vladimirov, A., Ellison, D. C., \& Bykov, A.
2006, \apj, 652, 1246

\bibitem[Webb et al.(1984)]{wdb84}
Webb, G. M., Drury, L. O'C., \& Biermann, P.
1984, \aap, 137, 185

\bibitem[Zank et al.(2006)]{zank06}
Zank, G. P., Li, G., Florinski, V., Hu, Q., Lario, D., \& Smith, C. W.  
2006, J. of Geophys. Res., 111, 06108

\bibitem[Zirakashvili \& Aharonian(2007)]{za07}
Zirakashvili V. N., \& Aharonian F. A.
2007, \aap, 465, 695

\end{thebibliography}
\end{document}